\documentclass[a4paper,reqno]{article}
\usepackage{amsmath,amssymb,amstext}
\usepackage[body={15cm,20.5cm}]{geometry}
\usepackage{url}

\begin{document}

\begin{center}

{\Large SLE$_\kappa$: correlation functions in  the coefficient problem}\\

\vspace{5mm}

Igor \textsc{Loutsenko}

\vspace{5mm}

Institut de Math\'ematiques de Jussieu,\\
Universit\'e Paris Diderot\\
e-mail: loutsenko@math.jussieu.fr\\[1mm]
April 20, 2012\\

\vspace{5mm}

Abstract

\end{center}

\begin{quote}
We apply the method of correlation functions to the coefficient problem in stochastic geometry. In particular, we give a proof for some universal patterns conjectured by M.~Zinsmeister for the second moments of the Taylor coefficients
for special values of $\kappa$ in the whole-plane Schramm-Loewner evolution (SLE$_\kappa$). We propose to use multi-point correlation functions for the study of higher moments in coefficient problem. Generalizations related to the Levy-type processes are also considered. The exact integral means $\beta$-spectrum of this version of the whole-plane SLE$_\kappa$ is discussed.
\end{quote}

\section{The coefficient problem for the whole plane SLE$_\kappa$: main results}

The Loewner evolution first appeared as a way for attacking the Bieberbach conjecture for the Taylor coefficients of injective conformal mappings of the open disc to the plane (for review see eg \cite{Gong}) and since the beginning of the 20th century became an indispensable tool in related theory of complex variables. The years following Loewners' initial work also showed many applications of complex analytic methods to the study of problems involving fractal objects in the plane. In particular, the area of two-dimensional critical phenomena has enjoyed a breakthrough due to a radical development by O.~Schramm in stochastic Loewner evolution, as an approach to description of boundaries of critical clusters (see for a review e.g. \cite{G} and \cite{GL2}).

On the other hand it is interesting to return to the initial motivation of studying Loewner evolution and to review the coefficient problem for stochastic conformal mappings described by Loewner chains. In particular, an idea of considering statistical properties of the Taylor coefficients of conformal mappings for stochastic Loewner evolution driven by a wide class of Levy processes has been recently proposed by M. Zinmeister. Some moments of coefficients have been estimated and some universal patterns have been conjectured by Bertrand Duplantier, Thi Phuong Chi Nguyen, Thi Thuy Nga Nguyen  and Michel Zinsmeister \cite{DNNZ}. In the present article we prove the conjectures of the above authors as well as propose new approaches to the coefficient problem in stochastic geometry.\\

We start the paper with a brief description of the problem and fixing the notations.\\

Radial stochastic Schramm-Loewner Evolution with parameter $\kappa$ (see e.g.\cite{G}, \cite{GL})  describes dynamics of a slit domain in the $z$-plane that can be represented by growth of a random planar curve $\Gamma=\Gamma(t)$ starting from a point on a unit circle $|z|=1$ at $t=0$. The principal idea of the theory of SLE is that the growth of the curve can be given by the time dependent conformal mapping $z=F(w,t)$ which satisfies the Loewner equation with stochastic driving:
$$
\frac{\partial F(w,t)}{\partial t}=w\frac{\partial F(w,t)}{\partial w}\frac{w+e^{\mathrm{i} \tilde B(t)}}{w-e^{\mathrm{i} \tilde B(t)}}, \quad  t\ge 0, \quad F(w,0)=w,
$$
where $\tilde B(t)$ denotes the Brownian motion with ``temperature'' $\kappa$:
$$
\langle (\tilde B(t)-\tilde B(t'))^2 \rangle =\kappa |t-t'|.
$$
Everywhere through the article $\langle\,\,\rangle$ denotes expectation.

One can consider either the ``exterior'' problem, where the complement of the unit disc $\mathbb{D}_+=\{w: |w|>1\}$ in the $w$-plane is mapped by $F=F_+$ to the ``exterior" slit domain $\mathbb{D}_+\backslash\Gamma(t)$ in the $z$-plane (i.e. the curve starting from the unit circle is growing in the exterior of the unit disc)
$$
F_+(w,t)=e^t\left(w+\sum_{i=0}^\infty\frac{F_i^+(t)}{w^i}\right), \quad |w|>1,
$$
or the ``interior'' problem, where the curve starting from the unit circle is growing in the interior of the unit disc. In this problem the unit disc in the $w$-plane $\mathbb{D}_-=\{w: |w|<1\}$ is mapped by $F=F_-$ to the ``interior" slit domain $\mathbb{D}_-\backslash\Gamma(t)$ in the $z$-plane
$$
F_-(w,t)=e^{-t}\left(w+\sum_{i=2}^\infty F_i^-(t)w^i\right), \quad |w|<1.
$$
Since SLE$_\kappa$ is a conformally invariant stochastic process \cite{G}, \cite{GL}, the exterior and interior problems are related by the inversion $F_-(w,t)=1/F_+(1/w,t)$.

In the present article we study the whole-plane SLE, which is an infinite-time limit of the radial SLE: There are two versions of the whole-plane SLE$_\kappa$:
\begin{equation}
\frac{\partial \mathcal{F}_\pm(w,t)}{\partial t}=\pm w\frac{\partial \mathcal{F}_\pm(w,t)}{\partial w}\frac{w+e^{\mathrm{i} B(t)}}{w-e^{\mathrm{i} B (t)}}, \quad \mathcal{F}_\pm(w,t)=e^{t}\left(w+\sum_{i=1\mp 1}^\infty \mathcal{F}_i^\pm(t)w^{\mp i}\right) .
\label{WholeSLE_fixed}
\end{equation}

The first version is the "interior" whole plane SLE$_\kappa$ which can be viewed as the limit of the interior problem
\begin{equation}
\mathcal{F}_-(w,t)=\lim_{T\to\infty}e^TF_-(w,T-t), \quad |w|<1, \quad B(t)=\tilde B(T-t)
\label{T_limit_SLE_int}
\end{equation}
describing the growth process in an infinite slit domain by ``erasing'' in time a curve/slit that starts at some point on the plane and goes to infinity. This article mainly deals with this version of the whole-plane SLE$_\kappa$.

The second version of the whole-plane SLE$_\kappa$ can be viewed as the limit of the exterior problem
\begin{equation}
\mathcal{F}_+(w,t)=\lim_{T\to\infty}e^{-T}F_+(w,T+t), \quad |w|>1,  \quad B(t)=\tilde B(T+t)
\label{T_limit_SLE_ext}
\end{equation}
Note that this is the bounded-map version of the whole-plane SLE$_\kappa$. Its exact multi-fractal spectrum has been recently described in \cite{BS}.

In the work \cite{DNNZ}, the coefficient problem (see e.g.~\cite{GL,Gong}) has been revisited in the framework of the SLE$_\kappa$: The authors of \cite{DNNZ} performed calculations of expectation values of squares of absolute values of several first Taylor coefficients $\langle |\mathcal{F}^-_i|^2\rangle$ (see Eq. (\ref{WholeSLE_fixed})) for the interior whole-plane stochastic Loewner evolution driven by Levy processes. They have observed some universal patterns, and in particular for the SLE$_\kappa$ with $\kappa=6$ and $\kappa=2$ they found that
$$
\kappa=2, \quad \langle |\mathcal{F}^-_n|^2\rangle =n
$$
$$
\kappa=6, \quad \langle |\mathcal{F}^-_n|^2\rangle =1
$$
In the present article we prove the above patterns (see Section 3) using technique of correlation functions, which is introduced in the next section. The multi-point correlation functions are considered in the context of evaluation of higher moments of expectations of the Taylor coefficients. This approach is also generalized to the case of the Loewner evolution driven by Levy processes.

Asymptotic behavior of the moments of coefficients, its relationship with orthogonal polynomials and multi-fractal spectrum is discussed in the last section.

\section{Correlation functions of SLE$_\kappa$}

One can find moments of Taylor coefficients through evaluation of moments of derivatives of conformal mappings. It turns out that moments of derivatives satisfy linear partial differential equations of the second order. To derive such equations it is convenient to change variable $w\to we^{\mathrm{i} B(t)}$ and define the conformal transformations  $z=f_\pm(w,t)$ in the ``rotating frame'', in which the immovable point $w=1$ on the unit circle is mapped to the moving tip of the curve:
\begin{equation}
f_\pm(w,t)=\mathcal{F}_\pm(we^{\mathrm{i} B(t)},t)=e^{t+\mathrm{i} B(t)}\left (w+\sum_{j=1\mp 1}^\infty f^\pm_j(t)w^{\mp j} \right) .
\label{rotation}
\end{equation}
We remind that the unbounded mapping $f_-$, corresponding to the interior whole-plane SLE$_\kappa$ (\ref{WholeSLE_fixed}), (\ref{T_limit_SLE_int}), sends the unit disc in the $w$-plane to the complement of a curve that starts at some point and goes to infinity in the $z$-plane. On the other hand, the bounded mapping $f_+$, corresponding to the exterior whole-plane SLE$_\kappa$ (\ref{WholeSLE_fixed}), (\ref{T_limit_SLE_ext}), sends the complement of the unit disc in the $w$-plane to the complement of a bounded curve that starts and ends at some finite points in the $z$-plane.

We now define the following functions
\begin{equation}
\rho_\pm(w,\bar w| q; \kappa)=e^{-qt}\langle (f'_\pm(w,t) \bar f'_\pm(\bar w,t))^{q/2}\rangle=\lim_{T\to\infty}e^{\mp qT}\langle (F'_\pm(we^{\mathrm{i} \tilde B(T)},T) \bar F'_\pm(\bar we^{-\mathrm{i} \tilde B(T)},T))^{q/2}\rangle,
\label{correlator}
\end{equation}
where the prime denotes derivative wrt $w$ or $\bar w$. In general, we do not suppose that the second argument $\bar w$ of the above function is a complex conjugate of the first one $w$, so we call it the ``$2$-point'' correlation function.

To estimate $\rho$ one can use the following differential equation
\begin{equation}
\mathcal{L}[\rho_\pm](w,\bar w|q;\kappa)=\sigma q \rho_\pm(w,\bar w|q;\kappa),
\label{Lrho}
\end{equation}
$$
\mathcal{L}=-\frac{\kappa}{2}\left(w\frac{\partial}{\partial w}-\bar w\frac{\partial}{\partial \bar w}\right)^2+\frac{w+1}{w-1}w\frac{\partial}{\partial w}+\frac{\bar w+1}{\bar w-1}\bar w\frac{\partial}{\partial \bar w}-\frac{q}{(w-1)^2}-\frac{q}{(\bar w-1)^2}+q
$$
where $\sigma=-1$ for the interior problem and $\sigma=1$ for the exterior problem respectively.

Below, we present a simple derivation of equation (\ref{Lrho}) which relies on procedure first introduced by Hastings \cite{H} (for a derivation using formal Ito calculus one can apply e.g. approach presented in \cite{BS}).

We start with the exterior whole-plane SLE$_\kappa$. From (\ref{correlator}) it follows that
\begin{equation}
\frac{\partial}{\partial t}\langle g \rangle=q \langle g \rangle, \quad g=(f'_+(w,t)\bar f'_+(\bar w,t))^{q/2}
\label{t}
\end{equation}
Using the fact that for the Loewner chain (\ref{WholeSLE_fixed}) we have the composition rule $\mathcal{F}_+(w,t+\delta t)=\mathcal{F}_+\left(\delta \mathcal{F}(w,\delta t),t\right)$, where $\delta \mathcal{F}$ itself satisfies the Loewner chain equation
\begin{equation}
\frac{\partial \delta \mathcal{F}(w,\delta t)}{\partial \delta t}=w\frac{\partial \delta \mathcal{F}(w,\delta t)}{\partial w}\frac{w+e^{\mathrm{i} B(t+\delta t)}}{w-e^{\mathrm{i} B (t+\delta t)}}, \quad \delta \mathcal{F}(w,\delta t=0)=w,
\label{delta F}
\end{equation}
in the ``rotating frame" (\ref{rotation}) we get
\begin{equation}
f_+(w, t+\delta t)=f_+\left(e^{-\mathrm{i} B(t)}\delta \mathcal{F}\left(we^{\mathrm{i} B(t+\delta t)},\delta t\right), t\right).
\label{r_composition}
\end{equation}
Solving differential equation (\ref{delta F}) up to the first order in $\delta t$ (up to the second order in $\delta B=B(t+\delta t)-B(t)$) we obtain
$$
e^{-\mathrm{i} B(t)}\delta \mathcal{F}\left(we^{\mathrm{i} B(t+\delta t)},\delta t\right)=w+\delta \phi(w)+\dots, \quad \delta\phi(w)=w\frac{w+1}{w-1}\delta t+\mathrm{i}w\delta B-\frac{w}{2}(\delta B)^2.
$$
Then, with the help of (\ref{t}), (\ref{r_composition}) we conclude that in the first order in $\delta t$:
$$
g(w, \bar w; t+\delta t)=g_+\left(w+\delta\phi(w), \bar w+\delta \bar \phi(\bar w); t\right)\left(1+\delta\phi'(w)\right)^{q/2}\left(1+\delta\bar\phi'(\bar w)\right)^{q/2}+\dots
$$
Equating expectations of RHS and LHS of the above and taking into account that $<\delta B>=0$, $<(\delta B)^2>=\kappa\delta t$ we find the time derivative of $\langle g\rangle$
$$
\frac{\partial}{\partial t}\langle g \rangle=\mathcal{L}\left[\langle g \rangle\right],
$$
where differential operator $\mathcal{L}$ is given in (\ref{Lrho}). Finally, with the help of (\ref{correlator}) and (\ref{t}) we arrive at differential equation (\ref{Lrho}) with $\sigma=1$ for the exterior whole-plane SLE$_\kappa$.

Since, according to (\ref{T_limit_SLE_int}), we defined the interior whole-plane SLE as a reversed Markovian process, equation for the interior whole-plane SLE$_\kappa$ has to be derived in the reversed time. This derivation coincides with that of the exterior case except the factor $e^{qt}$ must replace $e^{-qt}$ in (\ref{correlator}), and so we arrive at equation (\ref{Lrho}) with $\sigma=-1$. \\

\section{Interior Whole Plane SLE$_\kappa$}

According to (\ref{WholeSLE_fixed}) and (\ref{rotation}), expectations of squares of absolute values of Taylor coefficients are the same in both ``fixed" and ``rotating" frames $ \langle |f^-_j|^2 \rangle = \langle |\mathcal{F}_j^-|^2 \rangle $ and by the definition of two-point correlation function (\ref{correlator})
$$
\langle |\mathcal{F}_j^-|^2 \rangle=\rho^-_{j,j}(2,\kappa),
$$
where $ij\rho^-_{i,j}(q, \kappa)$ are the coefficients of the Taylor expansion of $\rho_-(w, \bar w| q; \kappa)$:
\begin{equation}
\rho_-(w, \bar w| q; \kappa)=\sum_{i=1}^\infty \sum_{j=1}^\infty ij \rho^-_{i,j}(q, \kappa)w^{i-1} \bar w^{j-1}, \quad \rho^-_{1,1}=1.
\label{whole_int}
\end{equation}

Substituting (\ref{whole_int}) into (\ref{Lrho}) we get the following recurrence relation for $\rho_{i,j}$:
\begin{equation}
\sum_{n=0}^2\sum_{k=0}^2 ij C_{i,j}^{n,k}\rho^-_{i-n,j-k}=0,  \quad \rho_{1,1}^-=1, \quad \rho_{i<0,j}^-=\rho_{i,j<0}^-=0,
\label{r_whole}
\end{equation}
$$
C_{i,j}^{0,0}=-\left(\frac{\kappa}{2}(i-j)^2+i+j-2\right), \quad
C_{i,j}^{1,1}=-4\left(\frac{\kappa}{2}(i-j)^2-2q\right), \quad
C_{i,j}^{2,2}=-\left(\frac{\kappa}{2}(i-j)^2-i-j+6-2q\right)
$$
$$
C_{i,j}^{0,1}=2\left(\frac{\kappa}{2}(j-i-1)^2+i-1-q\right), \quad C_{i,j}^{1,0}=2\left(\frac{\kappa}{2}(i-j-1)^2+j-1-q\right)
$$
$$
C_{i,j}^{0,2}=-\left( \frac{\kappa}{2}(j-i-2)^2+i-j+2-q\right), \quad
C_{i,j}^{2,0}=-\left( \frac{\kappa}{2}(i-j-2)^2+j-i+2-q\right),
$$
$$
C_{i,j}^{1,2}=2\left(\frac{\kappa}{2}(i-j+1)^2+3-j-2q\right), \quad C_{i,j}^{2,1}=2\left(\frac{\kappa}{2}(j-i+1)^2+3-i-2q\right)
$$
From this recurrence relation any $\rho^-_{i,j}(q,\kappa)$ can be found in a consecutive manner: We express $\rho^-_{i,j}$ as a linear combination of 8 coefficients $\rho^-_{i,j-1}$, $\rho^-_{i,j-2}$, $\rho^-_{i-1,j}$, $\rho^-_{i-1,j-1}$, $\rho^-_{i-1,j-2}$, $\rho^-_{i-2,j}$, $\rho^-_{i-2,j-1}$, $\rho^-_{i-2,j-2}$. At the first step we find $\rho^-_{1,2}$, then $\rho^-_{2,2}$ etc up to $\rho^-_{1,n}$. Repeating similar procedure for the second row $\rho^-_{2,j}, j=1..n$, then for the third row $\rho^-_{3,j}$... up to the $n$-th row $\rho^-_{n,j}, j=1..n$, we get all $\rho^-_{1..n, 1..n}$.

For example
$$
\rho^-_{2,2}(q,\kappa)=\frac{2q^2}{2+\kappa}
$$
$$
\rho^-_{3,3}(q,\kappa)=\frac{q^2}{36}\frac{9\kappa^2+8(14q+7+16q^2)\kappa+12(4q+1)^2}{(2+\kappa)(1+\kappa)(6+\kappa)}
$$
$$
\rho^-_{4,4}(q,\kappa)=\frac{q^2}{72}\left(240(2+3q+4q^2)^2+16\kappa^5+8(744q^4+340+1152q+2363q^2+1572q^3)\kappa^2\right.
$$
$$
+8(701q^2+378q^3+414q+192+144q^4)\kappa^3+32(635q^2+56+498q^3+272q^4+258q)\kappa
$$
$$
\left.+(204q+243q^2+284)\kappa^4\right)/
\left((2+\kappa)^2(1+\kappa)(6+\kappa)(2+3\kappa)(10+\kappa)\right)
$$
etc, whose particular $q=2$ case coincides with the results of computer experiments given in \cite{DNNZ}.\\

There are two particular values of $\kappa$ for which the following special results hold \\

\noindent
{\bf Theorem 1}:
\begin{itemize}

\item For $\kappa=6$, \quad
\begin{equation}
\langle |\mathcal{F}_n^-|^2 \rangle=1.
\label{kappa=6}
\end{equation}

\item For $\kappa=2$, \quad
\begin{equation}
\langle |\mathcal{F}_n^-|^2 \rangle=n.
\label{kappa=2}
\end{equation}
\end{itemize}

{\it Remark}: These patterns were first observed from direct calculations of several first Taylor coefficients in \cite{DNNZ}. \\

We prove the above results through solution of recurrence relation (\ref{r_whole}):

\begin{itemize}

\item For $\kappa=6$, the matrix $\rho^-_{i,j}$ is tri-diagonal, i.e. $\rho^-_{i,j}=0,\quad$ if $\quad |i-j|>1 $.
Nonzero matrix elements are $\rho^-_{i,i}=1,\quad$ and $\quad \rho^-_{i,i-1}=\rho^-_{i-1,i}=-1/2$.
One verifies the first statement of the theorem by direct substitution of $q=2, \kappa=6$, and the above matrix elements into (\ref{r_whole}).

\item For $\kappa=2$, the matrix $\rho^-_{i,j}$ is five-diagonal , i.e. $ \rho^-_{i,j}=0, \quad$ if $\quad |i-j|>2 $.
Nonzero matrix elements are $\rho^-_{i,i}=i, \quad \rho^-_{i,i-1}=\rho^-_{i-1,i}=(1-2i)/3, \quad \rho^-_{i,i-2}=\rho^-_{i-2,i}=(i-1)/6 $.
One verifies the second statement of the theorem by direct substitution of $q=2, \kappa=2$, and the above matrix elements into (\ref{r_whole}).
\end{itemize}

\noindent{\it Remark}\/:
An alternative proof can be given using the explicit solutions of (\ref{Lrho}) for $q=2$ and $\kappa=6, \kappa=2$
\begin{equation}
\rho_-(w, \bar w|q=2; \kappa=6)=\frac{(1-w)(1-\bar w)}{(1-w \bar w)^3}, \quad
\rho_-(w, \bar w|q=2; \kappa=2)=\frac{(1-w)^2(1-\bar w)^2}{(1-w \bar w)^4}
\label{kappa_6_2}
\end{equation}
It is interesting to note that equation (\ref{Lrho}) has one-parametric class of the explicit solutions
\begin{equation}
\rho_-\left(w, \bar w| q=\frac{(2+\kappa)(6+\kappa)}{8\kappa}; \kappa\right)=\frac{\left((1-w)(1-\bar w)\right)^{\frac{6+\kappa}{2\kappa}}}{\left(1-w \bar w\right)^{\frac{(6+\kappa)^2}{8\kappa}}}
\label{qkappa}
\end{equation}
which comprises the two above special cases.

\section{Exterior Whole-Plane SLE$_\kappa$}

In the exterior problem one considers the following expansion of $\rho_+$
\begin{equation}
\rho_+(w, \bar w| q; \kappa)=\sum_{i=-1}^\infty \sum_{j=-1}^\infty \frac{ij\rho_{i,j}^+(q, \kappa)}{w^{i+1}\bar w^{j+1}}, \quad \rho_{-1,-1}^+=1
\label{whole_ext}
\end{equation}
with the recurrence relation for $\rho_{i,j}^+$ having the form
$$
\sum_{n=0}^2\sum_{k=0}^2 ij C_{i,j}^{n,k}\rho^+_{i-n,j-k}=0, \quad \rho_{-1,-1}^+=1, \quad \rho^+_{i<-1,j}=\rho^+_{i,j<-1}=0,
$$
$$
C_{i,j}^{0,0}=-\left(\frac{\kappa}{2}(i-j)^2+i+j+2\right), \quad
C_{i,j}^{1,1}=-2\kappa(i-j)^2, \quad
C_{i,j}^{2,2}=-\left(\frac{\kappa}{2}(i-j)^2-i-j+2+2q\right)
$$
$$
C_{i,j}^{0,1}=2\left(\frac{\kappa}{2}(j-i-1)^2+i+1\right), \quad C_{i,j}^{1,0}=2\left(\frac{\kappa}{2}(i-j-1)^2+j+1\right)
$$
$$
C_{i,j}^{0,2}=-\left( \frac{\kappa}{2}(j-i-2)^2+i-j+2+q\right), \quad
C_{i,j}^{2,0}=-\left( \frac{\kappa}{2}(i-j-2)^2+j-i+2+q\right),
$$
$$
C_{i,j}^{1,2}=2\left(\frac{\kappa}{2}(i-j+1)^2-j+1+q\right), \quad C_{i,j}^{2,1}=2\left(\frac{\kappa}{2}(j-i+1)^2-i+1+q\right)
$$
Similarly to the interior case, any $\rho^+_{i,j}$ can be found recursively. Here we list several first $\langle|\mathcal{F}^+_i|^2\rangle$:
$$
\langle|\mathcal{F}_1^+|^2\rangle=\frac{1}{\kappa+1}
$$
$$
\langle|\mathcal{F}^+_2|^2\rangle=\frac{8\kappa(6+\kappa)}{9(\kappa+1)(3\kappa+2)(\kappa+10)}
$$
$$
\langle|\mathcal{F}^+_3|^2\rangle=\frac{\kappa(6+\kappa)(27\kappa^3+446\kappa^2+1300\kappa+264)}{36(\kappa+1)(\kappa+3)(3\kappa+2)(2\kappa+1)(\kappa+10)(\kappa+14)}
$$
etc.

\section{Multi-point correlation functions}

To estimate expectations of higher degrees of the Taylor coefficients, one needs to introduce multi-point correlation functions. For instance, the 4-point correlation function
$$
\rho_\pm(w_1, w_2, \bar w_1, \bar w_2| q_1, q_2; \kappa)=e^{-(q_1+q_2)t}\langle (f'_\pm(w_1,t) \bar f'_\pm(\bar w_1,t))^{q_1/2} (f'_\pm(w_2,t) \bar f'_\pm(\bar w_2,t))^{q_2/2} \rangle
$$
allows to estimate
$$
\langle \mathcal{F}_i^\pm\mathcal{F}_j^\pm \mathcal{\bar F}_l^\pm \mathcal{\bar F}_n^\pm \rangle, \quad i+j=l+n
$$
from the series expansion of $\rho_\pm(w_1, w_2, \bar w_1, \bar w_2|2, 2; \kappa)$. Similarly to the 2-point case the expansion coefficients can be found by solution of a PDE for the $2n$-point correlation function. Namely, the $2n$-point correlation function
\begin{align*}
&\rho_\pm(w_1, \dots,w_n, \bar w_1, \dots, \bar w_n| q_1, q_2, \dots,q_n; \kappa)\\
&\quad=e^{-(q_1+q_2+\dots,q_n)t}\langle (f'_\pm(w_1,t) \bar f'_\pm(\bar w_1,t))^{q_1/2}  \cdots (f'_\pm(w_n,t) \bar f'_\pm(\bar w_n,t))^{q_n/2}\rangle
\end{align*}
satisfies the following second-order linear PDE
\begin{equation}
\mathcal{L}[\rho_\pm]=\sigma \left(\sum_{i=1}^n q_i \right)\rho_\pm,
\label{L_2n}
\end{equation}
\begin{align*}
\mathcal{L}
&=\frac{\kappa}{2}\left(
\sum_{1\le i<j\le n}\left(w_i\frac{\partial}{\partial w_i}- w_j\frac{\partial}{\partial w_j}\right)^2
+\sum_{1\le i<j\le n}\left(\bar w_i\frac{\partial}{\partial \bar w_i}-\bar w_j\frac{\partial}{\partial \bar w_j}\right)^2-\sum_{i=1}^n\sum_{j=1}^n\left(w_i\frac{\partial}{\partial w_i}-\bar w_j\frac{\partial}{\partial \bar w_j}\right)^2\right)\\
&\quad
+\sum_{i=1}^n\left(\frac{w_i+1}{w_i-1}w_i\frac{\partial}{\partial w_i}+\frac{\bar w_i+1}{\bar w_i-1}\bar w_i\frac{\partial}{\partial \bar w_i}\right)
-\sum_{i=1}^n q_i\left(\frac{1}{(w_i-1)^2}+\frac{1}{(\bar w_i-1)^2}-1\right)
\end{align*}
where $\sigma=1$ for exterior problem and $\sigma=-1$ for interior problem respectively. \\

Similarly to the $2$-point case (\ref{correlator}), any coefficient $\rho_{\mathbf{i};\mathbf{j}}^-:=\rho_{i_1, i_2, \dots,i_n; j_1, j_2, \dots,j_n}^-$ of the Taylor expansion of the $2n$-point function
\[
\rho_-
=\sum_{i_1=1}^\infty\dots\sum_{i_n=1}^\infty\sum_{j_1=1}^\infty\dots\sum_{j_n=1}^\infty i_1i_2\cdots i_n j_1j_2\cdots j_n\rho_{\mathbf{i};\mathbf{j}}^-w_1^{i_1-1}\cdots w_n^{i_n-1}\bar w_1^{j_1-1}\cdots \bar w_n^{j_n-1},
\]
can be found in a consecutive manner using the ``$2n$-dimensional'' recursion relation
$$
\sum_{l_1=0}^2\dots\sum_{l_n=0}^2\sum_{k_1=0}^2\dots\sum_{k_n=0}^2i_1i_2\cdots i_n j_1j_2\cdots j_n C^{\mathbf{l};\mathbf{k}}_{\mathbf{i};\mathbf{j}}\rho^-_{\mathbf{i-l};\mathbf{j-k}}=0,
$$
where expressions for $C^{\mathbf{l};\mathbf{k}}_{\mathbf{i};\mathbf{j}}$ can be explicitly found by substitution of the above Taylor expansion of $\rho_-$ into (\ref{L_2n}).

\vspace{5mm}

\section{Non-Brownian processes}

The above considerations generalize straightforwardly to the case of stochastic Loewner evolution driven by Levy processes. In this case one considers an evolution of the type (\ref{WholeSLE_fixed}) with some Levy process $L(t)$ in place of the Brownian motion $B(t)$. Following Hastings' approach \cite{H}, one arrives to the analog of equation (\ref{Lrho}) for the two-point correlation function
\begin{equation}
-\hat\eta[\rho](w,\bar w)+\frac{w+1}{w-1}w\frac{\partial\rho(w, \bar w)}{\partial w}+\frac{\bar w+1}{\bar w-1}\bar w\frac{\partial\rho(w,\bar w)}{\partial \bar w}-q\left(\frac{1}{(w-1)^2}+\frac{1}{(\bar w-1)^2}-1+\sigma\right)\rho(w, \bar w)=0,
\label{eta_rho}
\end{equation}
where
$$
\hat\eta[\rho](w,\bar w)=-\lim_{t\to 0}\frac{\langle\rho\left(e^{\mathrm{i}L(t)} w, e^{-\mathrm{i}L(t)} \bar w\right)-\rho(w, \bar w)\rangle}{t}, \quad L(t=0)=0.
$$
Equations for multi-point correlation functions can be also derived in a similar manner.

As suggested in \cite{DNNZ} one can consider the coefficient problem for Levy processes with characteristic functions of the following type:
$$
\langle e^{\mathrm{i}\tau L(t)}\rangle=e^{-t\eta(\tau)}, \quad \eta(n)=\bar\eta(n)=\eta(-n).
$$
For these processes, in the case of the interior problem, we have
$$
\hat\eta[\rho](w,\bar w)=\eta\left(w\frac{\partial}{\partial w}-\bar w\frac{\partial}{\partial\bar w}\right)[\rho](w,\bar w)=\sum_{i=1}^\infty\sum_{j=1}^\infty ij\rho_{i,j}\eta_{i-j}w^{i-1}w^{j-1}.
$$
where $\eta_i:=\eta(i)$.

By analogy with the Brownian case, the expansion coefficients $\rho_{i,j}^-$ can be obtained from the recurrence relation of the type (\ref{r_whole}) with
$$
C_{i,j}^{0,0}=-\eta_{i-j}-i-j+2 , \quad
C_{i,j}^{1,1}=-4(\eta_{i-j}-2q) , \quad
C_{i,j}^{2,2}=-\eta_{i-j}+i+j-6+2q ,
$$
$$
C_{i,j}^{0,1}=2\left(\eta_{i-j+1}+i-1-q\right), \quad C_{i,j}^{1,0}=2\left(\eta_{i-j-1}+j-1-q\right) ,
$$
$$
C_{i,j}^{0,2}=-\eta_{i-j+2}+j-i-2+q, \quad C_{i,j}^{2,0}=-\eta_{i-j-2}+i-j-2+q ,
$$
$$
C_{i,j}^{1,2}=2\left(\eta_{i-j+1}+3-j-2q\right), \quad C_{i,j}^{2,1}=2\left(\eta_{i-j-1}+3-i-2q\right) .
$$

As conjectured in \cite{DNNZ} (by calculations of $\langle |\mathcal{F}_n|^2\rangle$ for $n\le 20$), results (\ref{kappa=6}), (\ref{kappa=2}) also hold for Levy processes with $\eta_1=1$ and $\eta_1=3$, that correspond to $\kappa=2, \kappa= 6$ in the SLE$_\kappa$ case.

The case $\eta_1=3$ can be easily proved either by repeating arguments of Theorem 1 that now use the above recurrence coefficients or by explicit solution of equation (\ref{eta_rho}): The $\eta_1=3$ solution of (\ref{eta_rho}) coincides with the $q=2,\kappa=6$ solution (\ref{kappa_6_2}) of equation (\ref{Lrho}).\\

The case $\eta_1=1$ seems to be more involved. Note, however, that one can also repeat arguments of Theorem 1 in the $\eta_2=4$ subcase of this case, since corresponding solution of (\ref{eta_rho}) coincides with the $q=2,\kappa=2$ solution of equation (\ref{Lrho}).

\section{Orthogonal polynomials and multi-fractal spectrum}

In conclusion, we would like to present some observations related to the 2-dimensional recurrence relation (\ref{r_whole}) which seems to be an interesting structure:

In the interior case, the first line $\rho^-_{1,j}$ of matrix $\rho^-_{i,j}$ corresponds to the function $\rho_-(0,\bar w|q;\kappa)$
\begin{equation}
\rho_-(0,\bar w|q;\kappa)=\sum_{j=1}^\infty j\rho^-_{1,j}\bar w^{j-1}
\label{Taylor}
\end{equation}
that satisfies the $w=0$ reduction of equation (\ref{Lrho}). The reduced equation is amenable, by a gauge transform, to the Gauss hypergeometric equation. To find $\rho_{1,j}$ one can either choose an appropriate solution of this hypergeometric equation or consider the three-term recurrence relation for $\rho^-_{1,j}$ following directly from (\ref{r_whole}). The latter recurrence relation is related to Dual Hahn polynomials \cite{KLS}) and the expansion (\ref{Taylor}) truncates at $j=N+1$ (i.e. $\rho^-_{1,j}=0, j>N+1$) if
\begin{equation}
q=\frac{Nn(2N-n+1)}{N^2+n^2-n}, \quad \kappa=2\frac{q+N}{N^2}=2\frac{2n+N}{N^2+n^2-n}, \quad n\le N, \quad N,n=1,2,3 \dots
\label{spectrum}
\end{equation}
Furthermore, since for the above values of parameters, $C^{0,2}_{i,i+N+2}=0$ in the recurrence relation (\ref{r_whole}), the whole matrix $\rho^-_{i,j}$ becomes $2N+1$-diagonal, i.e. $\rho^-_{i,j}=0$ if $|i-j|>N$ when (\ref{spectrum}) holds.\\

\noindent{\it Remark}\/: The cases $q=2, \kappa=6$ and $q=2, \kappa=2$ of the Theorem 1 correspond to $N=1, n=1$ and $N=2, n=2$ in (\ref{spectrum}).\\


It is worth to note that the orthogonal polynomials also appear in asymptotic expansion of $\rho^-(w, \bar w)$: In more details, the multi-fractal spectrum of $SLE_\kappa$ is determined by asymptotic behavior of diagonal coefficients $\rho^-_{i,i}(q,\kappa)\to{\cal C}(q,\kappa)i^{\beta(q,\kappa)-3}$ in the $i\to\infty$ limit,
where $\beta(q,\kappa)$ is an integral ``means $\beta$-spectrum'':
$$
\beta=\lim_{\epsilon\to0}\frac{\log \int_0^{2\pi}\rho_-\left(e^{-\epsilon+\i\theta},e^{-\epsilon-\i\theta}|q;\kappa\right)d\theta}{-\log \epsilon}.
$$
Using the Ansatz
$$
\rho^-_{i,i+l}\to f_l i^{\beta-3} , \quad i\to \infty,
$$
which represents the highest term in the asymptotic expansion $\rho^-_{i,i+l}=f_l i^{\beta-3}+f_l^{(1)} i^{\beta-4}+\dots$, from the recurrence relation (\ref{r_whole}) in the leading order in $i$ we get
$$
r_{l+1}+r_{l-1}-2r_l=0, \quad r_l=R[f]_l-\beta f_l,
$$
where $R$ is the three-diagonal difference operator
\begin{equation}
R[f]_l=\frac{1}{2}\left(a_{l+1}f_{l+1}+b_lf_l+c_{l-1}f_{l-1}\right),
\label{Delta_3}
\end{equation}
$$
a_l=\frac{\kappa l^2}{2}+l-q, \quad c_l=a_{-l}=\frac{\kappa l^2}{2}-l-q, \quad b_l=-a_l-c_l+2q=-\kappa l^2+4q .
$$
In the case when
\begin{equation}
f_l\to 0, \quad l\to\pm\infty,
\label{f_nlarge}
\end{equation}
all $r_l$ vanish and
\begin{equation}
R[f]_l=\beta f_l.
\label{Delta_3_Lambda}
\end{equation}
Thus, provided (\ref{f_nlarge}) holds, the value of the $\beta$-spectrum for given $\kappa$ and $q$ is an eigenvalue of the three-term difference operator $R$. \\

{\it Remark}: The above asymptotic analysis generalizes to the case of the Levy processes considered in the previous section with $\beta$ being an eigenvalue of the three diagonal difference operator of the type (\ref{Delta_3}), (\ref{Delta_3_Lambda}) with $ a_l=\eta_l+l-q$, $ c_l=a_{-l}$, $ b_l=-a_l-c_l+2q $.\\

In the case when parameters $q$ and $\kappa$ take values (\ref{spectrum}), i.e. in the case of the $2N+1$ band truncation of $\rho^-_{i,j}$, the condition (\ref{f_nlarge}) holds and $f_l$ satisfy equation (\ref{Delta_3_Lambda}), which is the three-term recurrence relation for the Dual Hahn polynomials \cite{KLS} with the following spectrum:
\begin{equation}
\beta=\frac{2N(n+6nN-3n^2-N)-(8nN-2n^2-N+2N^2)j+(2n+N)j^2}{2(N^2+n^2-n)},
\label{hahn_spectrum}
\end{equation}
where $j\in\{0,1,2,\dots,2N\}$. To choose a proper $j$, corresponding to a value of the $\beta$-spectrum for specified parameters, additional analysis is needed.

Note, that $n=N$ subset of (\ref{spectrum}) belongs to the one-parametric family $ q=\frac{(2+\kappa)(6+\kappa)}{8\kappa} $ of solutions (\ref{qkappa}) (including the cases of Theorem 1) . Using these exact solutions one can verify that for this subset $j=0$ in (\ref{hahn_spectrum}), which corresponds to the highest eigenvalue of the difference operator $R$. With the help of (\ref{spectrum}) and (\ref{hahn_spectrum}) one can rewrite expression for this eigenvalue in terms of $q$ and $\kappa$:
\begin{equation}
\beta=3q-1-\frac{q\kappa}{1+\sqrt{1+2q\kappa}}.
\label{beta-spectrum}
\end{equation}
It is easy to see that the above expression for the $\beta$-spectrum is valid for the whole one-parametric family of solutions (\ref{qkappa}), i.e. for all real non-negative $\kappa$ and $ q=\frac{(2+\kappa)(6+\kappa)}{8\kappa} $. General consideration from the theory of harmonic measure show that (\ref{beta-spectrum}) equals the value of $\beta$-spectrum for all elements of set (\ref{spectrum}) \cite{LY}. Therefore, one may conjecture that equation (\ref{beta-spectrum}) gives expression for the $\beta$-spectrum of the considered version of the whole plane SLE$_\kappa$  in a certain domain of the parametric $(q,\kappa)$-plane that includes families (\ref{qkappa}) and (\ref{spectrum}). The latter is confirmed by numerical analysis \cite{LY}. \\

In conclusion: It is interesting to understand what is the particular significance of the considered $2N+1$-diagonal structures and whether the whole two-dimensional recurrence relation (\ref{r_whole}) is ``exactly solvable''.

\vspace{5mm}

\noindent
{\large\textbf{Acknowledgement}}\\

We would like to acknowledge help received from V.~Spiridonov and A.~Zhedanov. This work has been supported by the European Commission 7th framework IEF grants.


\vspace{5mm}

\begin{thebibliography}{99}

\bibitem{DNNZ}
Bertrand Duplantier, Thi Phuong Chi Nguyen, Thi Thuy Nga Nguyen, Michel Zinsmeister, \emph{Coefficient estimates for whole-plane SLE processes}, \url{http://hal.inria.fr/hal-00609774}, 2011.

\bibitem{BS} D. Beliaev, S. Smirnov,
\emph{Harmonic measure and SLE}, Commun. Math. Phys. 290, 577–595 (2009).

\bibitem{G}
Ilya A.~Gruzberg,
\emph{Stochastic geometry of critical curves, Schramm--Loewner evolutions and conformal field theory}, J.~Phys.~A: Math.~Gen., \textbf{39}, no.~41 (2006) 12601--12655.

\bibitem{Gong}
Sheng Gong,
\emph{The Bieberbach conjecture},
AMS/IP Studies in Advanced Mathematics, \textbf{12}, Amer. Math. Soc., Providence, RI, International Press, Cambridge, MA, 1999.

\bibitem{H}
Matthew B.~Hastings,
\emph{Exact Multifractal Spectra for Arbitrary Laplacian Random Walks},
Phys. Rev. Lett., \textbf{88} (2002) 055506.

\bibitem{KLS}
Roelof Koekoek, Peter A.~Lesky, Ren\'e F.~Swarttouw,
\emph{Hypergeometric orthogonal polynomials and their $q$-analogues}, Springer-Verlag, Berlin, 2010.

\bibitem{GL}
Gregory F.~Lawler,
\emph{Conformally invariant processes in the plane},
Mathematical Surveys and Monographs, \textbf{114}, Amer. Math. Soc., Providence, RI, 2005.

\bibitem{GL2}
Gregory F.~Lawler,
\emph{Conformal invariance and 2D statistical physics},
Bull. Amer. Math. Soc., \textbf{46} (2009) 35--54.

\bibitem{LY} I.~Loutsenko, O.~Yermolayeva, in preparation


\end{thebibliography}
\end{document}